\documentclass[prd,amsmath,amssymb,superscriptaddress,preprintnumbers,twocolumn,nofootinbib,10pt]{revtex4-1}
\pdfoutput=1
\usepackage{graphicx}
\usepackage{dcolumn}
\usepackage{bm}
\usepackage{amssymb}
\usepackage{latexsym}
\usepackage{booktabs}
\usepackage{amsmath}
\usepackage{multirow}
\usepackage{url}
\usepackage{footnote}
\usepackage{float}
\usepackage{threeparttable}
\usepackage[colorlinks=true, linkcolor=blue, citecolor=blue]{hyperref}
\usepackage[bottom]{footmisc}

\usepackage[normalem]{ulem}
\usepackage{color}
\usepackage{array}
\usepackage{enumerate}
\usepackage{adjustbox}

\usepackage{makecell}
\usepackage{diagbox}
\usepackage{epstopdf}
\usepackage{epsfig}
\usepackage{longtable}
\usepackage{supertabular}
\usepackage{algorithm}
\usepackage{pifont}
\usepackage{algorithmic}
\usepackage{changepage}
\usepackage{setspace}
\begin{document}

\title{Probing the sign-changeable interaction between dark energy and dark matter with DESI baryon acoustic oscillations and DES supernovae data} 

\author{Tian-Nuo Li}
\affiliation{Key Laboratory of Cosmology and Astrophysics of Liaoning Province, College of Sciences, Northeastern University, Shenyang 110819, China}

\author{Guo-Hong Du}
\affiliation{Key Laboratory of Cosmology and Astrophysics of Liaoning Province, College of Sciences, Northeastern University, Shenyang 110819, China}

\author{Yun-He Li}
\affiliation{Key Laboratory of Cosmology and Astrophysics of Liaoning Province, College of Sciences, Northeastern University, Shenyang 110819, China}

\author{Peng-Ju Wu}
\affiliation{School of Physics, Ningxia University, Yinchuan 750021, China}

\author{Shang-Jie Jin}
\affiliation{Key Laboratory of Cosmology and Astrophysics of Liaoning Province, College of Sciences, Northeastern University, Shenyang 110819, China}
\affiliation{Department of Physics, University of Western Australia, Perth WA 6009, Australia}

\author{Jing-Fei Zhang}
\affiliation{Key Laboratory of Cosmology and Astrophysics of Liaoning Province, College of Sciences, Northeastern University, Shenyang 110819, China}

\author{Xin Zhang}\thanks{Corresponding author}\email{zhangxin@mail.neu.edu.cn}
\affiliation{Key Laboratory of Cosmology and Astrophysics of Liaoning Province, College of Sciences, Northeastern University, Shenyang 110819, China}
\affiliation{MOE Key Laboratory of Data Analytics and Optimization for Smart Industry, Northeastern University, Shenyang 110819, China}
\affiliation{National Frontiers Science Center for Industrial Intelligence and Systems Optimization, Northeastern University, Shenyang 110819, China}

\begin{abstract}

There is a possibility of interaction between dark energy and dark matter, and it may undergo a sign change during the evolution of the universe. In this paper, we utilize the latest DESI baryon acoustic oscillation data, along with type Ia supernova data from DES and cosmic microwave background data from Planck and ACT, to constrain models of sign-changeable interaction. From our analysis, we observe that the coupling $\beta(z)$ crosses the non-interacting line $\beta(z) = 0$ and changes sign from positive to negative during cosmic evolution. Specifically, we find that the existence of sign-changeable interaction in the model with $Q = \beta(a)H_0\rho_{\rm de}$ is supported by the current data at the 4.1$\sigma$ level. Our findings indicate that the energy transfer is from dark matter to dark energy when dark matter dominates the universe, and from dark energy to dark matter when dark energy dominates, for the models with $Q \propto \rho_{\rm de}$. Furthermore, Bayesian evidence suggests that the $Q \propto \rho_{\rm de}$ models are moderately preferred over the $\Lambda$CDM model. The overall outcomes of this study clearly indicate that, based on current observational data, the sign-changeable interacting dark energy models are quite compelling and merit further attention.

\end{abstract}
\maketitle

\section{Introduction}
Over the past two decades, a major unresolved challenge in contemporary cosmology has been to understand the physical mechanism driving the universe's late-time accelerated expansion, first discovered through type Ia supernova (SN) observations \cite{SupernovaSearchTeam:1998fmf,SupernovaCosmologyProject:1998vns}. To explain the observed cosmic accelerated expansion, the concept of dark energy (DE), an exotic form of energy with negative pressure, has been introduced. DE accounts for around 68\% of the overall energy density of the universe and therefore governs the evolution of the current cosmos. In its simplest form, DE is assumed to be a cosmological constant $\Lambda$, with the equation of state (EoS) being $w = -1$. The cosmological constant, along with the cold dark matter (CDM), a key component for structure formation in the universe, plays a crucial role in setting up the standard cosmological model, known as the $\Lambda$CDM model.

The standard $\Lambda$CDM model, although it has achieved significant successes over the past decades, appears to fall short in providing a comprehensive explanation for the latest observations, such as the $S_8$ tension\footnote{Recent cosmic shear constraints from the Kilo-Degree Survey suggest that the $S_8$ tension has been significantly alleviated following improved redshift distribution estimation and calibration, with a discrepancy of only $0.73\sigma$ relative to the Planck results \cite{Wright:2025xka}.} \cite{DiValentino:2020vvd} and the $H_0$ tension \cite{Verde:2019ivm,Riess:2019qba,Riess:2021jrx}. Specifically, the $H_0$ tension is a discrepancy exceeding 5$\sigma$ between the cosmic microwave background (CMB) estimates of the Hubble constant (assuming a $\Lambda$CDM cosmology) \cite{Planck:2018vyg} and the SH0ES (Supernovae and $H_0$ for the EoS of DE) measurements \cite{Riess:2021jrx}. Recently, the $H_0$ tension has been extensively explored and discussed in the literature \cite{Bernal:2016gxb,Guo:2018ans,Vagnozzi:2019ezj,DiValentino:2021izs,Cai:2021wgv,Escudero:2022rbq,Zhao:2022yiv,James:2022dcx,Kamionkowski:2022pkx,Vagnozzi:2023nrq,Jin:2022qnj,Zhang:2023gye,Pierra:2023deu,Song:2022siz,Jin:2023sfc,Huang:2024gfw,Huang:2024erq,Zhang:2024rra}. Additionally, on the theoretical front, the $\Lambda$CDM model encounters issues like the ``fine-tuning" and ``cosmic coincidence" problems \cite{Sahni:1999gb,Bean:2005ru}. These issues may suggest the existence of new physics beyond the standard $\Lambda$CDM model. This has motivated our scientific community to build complex DE models, such as early dark energy model \cite{Doran:2006kp}, dynamical dark energy model \cite{Chevallier:2000qy}, holographic dark energy model \cite{Li:2004rb,Zhang:2005hs}, and interacting dark energy (IDE) model. The IDE model postulates a non-gravitational interaction between DE and dark matter (DM), enabling an exchange of energy between them. It has been found that the IDE models can not only help alleviate the coincidence problem \cite{Chimento:2003iea,Hu:2006ar,Dutta:2017kch}, $S_8$ tension \cite{Lucca:2021dxo}, and $H_0$ tension \cite{DiValentino:2017iww,Yang:2018euj,DiValentino:2019ffd,Gao:2021xnk}, but also help probe the fundamental nature of DE and DM.
In recent years, the IDE models have been extensively discussed \cite{Farrar:2003uw,Zhang:2005rj,Wang:2005jx,Setare:2006wh,Zhang:2007uh,Li:2009zs,Li:2010ak,Costa:2013sva,Li:2014cee,Costa:2016tpb,Zhang:2017ize,Pan:2019gop,Li:2019ajo,Feng:2019mym,Pan:2020zza,DiValentino:2020kpf,Zhang:2021yof,Qi:2021iic,Lucca:2021eqy,Wang:2022llq,Jin:2022tdf,Zhao:2022bpd,Teixeira:2023zjt,Yao:2023jau,Li:2023gtu,Forconi:2023hsj,Han:2023exn,Giare:2024ytc,Benisty:2024lmj,Feng:2024lzh}; see also refs.~\cite{Wang:2016lxa,Wang:2024vmw} for reviews.

Recently, the Dark Energy Spectroscopic Instrument (DESI) released the first-year data (i.e., 2024 Data Release 1), providing baryon acoustic oscillations (BAO) measurements from galaxy, quasar, and Lyman-$\alpha$ (Ly$\alpha$) forest tracers \cite{DESI:2024uvr,DESI:2024lzq}. The DESI BAO data, combined with CMB data from the Planck satellite and the Atacama Cosmology Telescope (ACT) as well as SN data from the Dark Energy Survey Year 5 (DESY5), have provided $3.9\sigma$ evidence for dynamical DE under the $w_0w_a$CDM model \cite{DESI:2024mwx}. Interestingly, \citet{Park:2024vrw} has also reported similar evidence for dynamical DE using the $w_0w_a$CDM model with other BAO data. The evidence for dynamical DE observed in the DESI BAO data has sparked extensive discussions regarding DE. Recently, the DESI BAO data along with other data like CMB and SN have been considered to constrain various aspects of cosmological physics (see, e.g., \cite{Giare:2024gpk,Dinda:2024ktd,Jiang:2024xnu,Jiang:2024viw,Escamilla:2024ahl,Sabogal:2024yha,Reboucas:2024smm,Pang:2024qyh,Giare:2024oil,Wang:2024tjd,Alestas:2024eic,Chan-GyungPark:2024brx,Specogna:2024euz,Wu:2024faw,Li:2024qus,Du:2024pai,Wang:2024rus,Fikri:2024klc,Toda:2024uff,Gao:2024ily,Giare:2024smz,Yang:2024kdo,RoyChoudhury:2024wri,Chan-GyungPark:2025cri,Li:2025eqh}). Furthermore, we specifically conducted a study using the DESI BAO data combined with the CMB and DESY5 SN data to constrain the IDE models \cite{Li:2024qso}.

In our previous work, the interaction term $Q$, which controls the energy transfer between DE and DM, was considered in the following four typical forms. Most of these forms are specifically constructed for mathematical simplicity, with the forms of $Q$ usually assumed to be proportional to the energy density of the dark sectors \cite{Amendola:1999qq,Billyard:2000bh}, such as $Q=\beta H\rho_{\rm c}$ or $Q=\beta H\rho_{\rm de}$. In addition, another perspective suggests that $Q$ should not involve the Hubble parameter $H$ because the local interaction should not depend on the global expansion of the universe \cite{Boehmer:2008av,Valiviita:2008iv,Clemson:2011an}. Thus, according to this perspective, another form of $Q$ is assumed, such as $Q=\beta H_0\rho_{\rm c}$ or $Q=\beta H_0\rho_{\rm de}$, where the inclusion of $H_0$ is merely for dimensional reasons. However, it is worth noting that, in these models, since the coupling parameter $\beta$ is a constant, the flow of energy between the dark sectors is assumed to be unidirectional. That is, throughout the period of energy exchange between the dark sectors, the energy transfer can occur either from DE to DM or from DM to DE. However, based on theoretical and observational grounds, there is evidence suggesting that the direction of energy flow may reverse in the late-time universe \cite{Wei:2010cs,Sun:2010vz,Forte:2013fua,Guo:2017deu,Arevalo:2019axj,Pan:2019jqh,Sobhanbabu:2021vzw,Halder:2024uao,Arevalo:2022sne,Mahata:2015lja,Chakraborty:2020yfe,Bahamonde:2017ize,Caldera-Cabral:2008yyo,Shahalam:2015sja,Bernardi:2016xmb,Zonunmawia:2017ofc,Hussain:2022dhp,Kritpetch:2024rgi}. This is an interesting question in the context of IDE scenarios and is referred to as a sign-changeable IDE model.

Cai and Su \cite{Cai:2009ht} were the first to explore the scenario of sign-changeable interaction. Instead of selecting a specific phenomenological interaction form, they implemented a piecewise approach, dividing the entire redshift range into several intervals and treating the coupling $\delta(z)$ (with $Q=3H\delta$ in their model) as a constant within each interval. Based on the fitting results of the observational data available at that time, they concluded that $\delta(z)$ is likely to cross the noninteracting line ($\delta=0$). However, the limitation of this work lies in the fact that the sign-change behavior is derived from the best-fit $\delta(z)$, making it inconclusive, as the errors in the fit results significantly undermine the reliability of the conclusion. In the context of a piecewise parametrization approach, the observational data are insufficient to constrain more than two parameters. 

To address this issue, Li and Zhang \cite{Li:2011ga} introduced a parameterization for the interaction term, expressed as $Q(a)=3\beta(a)H_0\rho_0$, where $\beta(a)$ represents a dimensionless coupling that evolves over the course of cosmic evolution, $a$ is the scale factor of the universe, $H_0$ is the Hubble constant, and $\rho_0=\rho_{\rm de0}+\rho_{\rm c0}$ is the present-day density of dark sectors. Since the evolution of the interaction term $Q$ is completely described by the running of the coupling parameter $\beta(a)$, this is called the ``running coupling'' scenario. The parametrized form of the coupling is given as
\begin{equation}\label{b}
\beta(a)=\beta_0a+\beta_{\rm e}(1-a),
\end{equation}
where $\beta_0$ and $\beta_{\rm e}$ are two dimensionless parameters. The coupling parameter $\beta(a)$ is governed by $\beta_0$ at the late-time universe and by $\beta_{\rm e}$ at the early-time universe. Thus, the full evolution of $\beta(a)$ is entirely determined by the two parameters, $\beta_0$ and $\beta_{\rm e}$, with Eq.~(\ref{b}) providing a continuous connection between the early-time and late-time behaviors. Subsequently, \citet{Guo:2017deu} used this approach to study six sign-changeable IDE models, utilizing the observational data available at the time. Their results showed that the coupling changes sign during cosmic evolution at approximately the $1\sigma$ confidence level. 

Interestingly, \citet{Escamilla:2023shf} used model-independent methods to study the interaction between DE and DM, also indicating a possible sign change in the direction of energy transfer. This provides sufficient motivation to incorporate a sign-changeable nature into the interaction terms and explore the resulting implications. 

In this work, we aim to investigate whether the interaction between DE and DM changes sign during cosmic evolution, using the DESI BAO, CMB, and DESY5 SN data. In addition, to achieve high generality, we investigate four different forms of the interaction term $Q$. Finally, we utilize Bayesian evidence to evaluate which form of the sign-changeable IDE model is preferred by the current observational data.

This work is organized as follows. In Sec.~\ref{sec2}, we briefly introduce the IDE models, as well as the cosmological data used in this work. In Sec.~\ref{sec3}, we report the constraint results and make some relevant discussions. The conclusion is given in Sec.~\ref{sec4}.

\section{Methodology and data}\label{sec2}

\subsection{Brief description of the IDE models}\label{sec2.1}

For the IDE model, the conservation laws for the energy-momentum tensor ($T_{\mu\nu}$) of DE and CDM are modified as 
\begin{equation}
  \label{eq:energyexchange} \nabla_\nu T^\nu_{\hphantom{j}\mu,\rm de} = -\nabla_\nu T^\nu_{\hphantom{j}\mu,\rm c}  = Q_\mu,
\end{equation} 
where $Q_\mu$ denotes the energy-momentum transfer vector. A specific $Q_{\mu}$ determines the energy transfer rate $Q$, its perturbation $\delta Q$ and the momentum transfer rate $f$.

In the background level, the energy conservation equation can be written as
\begin{align}\label{conservation1}
{\rho}_{\rm de}^{\prime } +3\mathcal{H}(1+w)\rho_{\rm de}= aQ,\\
{\rho}_{\rm c}^{\prime } +3 \mathcal{H} \rho_{\rm c}= -aQ,
\end{align}
where prime denotes differentiation with respect to the conformal time, $\rho_{\rm{de}}$ and $\rho_{\rm c}$ represent the energy densities of DE and CDM, $\mathcal{H}= a H$ is the conformal Hubble parameter, $a$ is the scale factor, $w$ is the EoS parameter of DE, and $Q$ denotes the interaction term describing the energy transfer rate between DE and CDM due to the interaction.

In this work, we only wish to extend the base $\Lambda$CDM model in a minimal way. Thus, we focus solely on the case where $w = -1$ to avoid introducing additional parameters, as also adopted in related studies \cite{Xu:2011qv,Chimento:2013rya,Wang:2014xca,SanchezG:2014snn,Guo:2017deu,Guo:2017hea,Wang:2021kxc}. It should be emphasized that setting $w = -1$ in a scenario with interaction between DE and CDM does not necessarily lead to a constant DE density over space and time. Actually, in this case, the DE density $\rho_{\rm de}$ is now a time-evolving quantity because ${\rho}_{\rm de}^{\prime } = aQ$ (see Eq.~(\ref{conservation1}), with $w=-1$). Furthermore, one can see that the conservation equations can be rewritten by introducing effective EoS for the DE and CDM as
\begin{align}\label{conservation2}
{\rho}_{\rm de}^{\prime }+3\mathcal{H}\left( 1+w_{\rm de}^{\mathrm{eff}}\right) \rho
_{\rm de} = 0, \\
{\rho}_{\rm c}^{\prime }+3\mathcal{H}\left( 1+w_{\rm c}^{\mathrm{eff}}\right) \rho
_{\rm c} = 0,  
\end{align}
where $w_{\rm de}^{\mathrm{eff}}$, $w_{\rm c}^{\mathrm{eff}}$ are defined as the
effective EoS parameters for DE and CDM with 
\begin{align}\label{effective equation}
w_{\rm de}^{\mathrm{eff}} = w-\frac{aQ}{3\mathcal{H}\rho _{\rm de}}, \\ 
w_{\rm c}^{\mathrm{eff}} = \frac{aQ}{3\mathcal{H}\rho _{\rm c}}.
\end{align}
We note that the effective EoS of DE admits multiple behaviors, determined by the specific interaction term $Q$.

For the interaction term $Q$, in the absence of a fundamental theory, we adopt a phenomenological approach in which the interaction term is commonly assumed to be proportional to the energy density of either DE or CDM \cite{Amendola:1999qq, Billyard:2000bh}. To balance the dimensions, it needs to be multiplied by a quantity with units of the inverse of time. The obvious choice is Hubble parameter, as it can provide an analytical solution for the conservation equations. Thus, the first two interaction terms take the following forms 
\begin{align}
\mbox{IDE1:} \quad & Q = \beta(a)H\rho_{\rm de}, \\
\mbox{IDE2:} \quad & Q = \beta(a)H\rho_{\rm c},
\end{align}
where $\beta(a)$ is the dimensionless time-variable coupling parameter of the interaction function. We assume that the coupling $\beta(a)$ is described by a constant $\beta_0$ at the late times, and determined by another constant $\beta_{\rm e}$ at the early times. For continuously connecting the early-time and late-time behaviors, we adopt the following two-parameter form for the coupling $\beta(a)=\beta_0a+\beta_{\rm e}(1-a)$. Note that the purpose of this parameterization is solely to investigate whether the interaction between DE and CDM undergoes a sign change during cosmic evolution. Furthermore, in the research area of IDE, there is another perspective that $Q$ should not involve Hubble parameter because the local interaction should not depend on the global expansion of the universe \cite{Valiviita:2008iv,Boehmer:2008av,Caldera-Cabral:2008yyo,He:2008si,Clemson:2011an}. Thus, the other two models that we introduce are given by the following forms
\begin{align}
\mbox{IDE3:} \quad & Q = \beta(a)H_{0}\rho_{\rm de}, \\
\mbox{IDE4:} \quad & Q = \beta(a)H_{0}\rho_{\rm c}.
\end{align}
Notice that the appearance of $H_0$ is solely for dimensional considerations.

In the linear perturbation level, Eq.~(\ref{eq:energyexchange}) is given by
\begin{widetext}
\begin{equation}
 {\delta\rho_I'} + 3\mathcal{H}({\delta \rho_I}+ {\delta p_I})+(\rho_I+p_I)(k{v}_I + 3 H_L') = a(\delta Q_I + AQ_I), \label{eqn:conservation1}
\end{equation}
\begin{equation}
[(\rho_I + p_I)(v_I - B)]' + 4\mathcal{H}(\rho_I + p_I)(v_I - B) - k \delta p_I + \frac{2}{3}k c_K p_I \Pi_I - k (\rho_I + p_I) A = a[Q_I(v - B) + f_I], \label{eqn:conservation2}
\end{equation}
\end{widetext}
where $I$ represents $\rm DE$ or $\rm CDM$, $\delta\rho_I$ is energy density perturbation, $\delta p_I$ is isotropic pressure perturbation, $A$, $B$, and $H_L$ are used to describe the scalar metric perturbations, $v_I$ is velocity perturbation, $\Pi_I$ is anisotropic stress perturbation, and $c_K = 1-3K/k^2$ with $K$ being the spatial curvature. Here, we assume a flat universe ($K = 0$). For the computation of metric perturbations, we adopt the synchronous gauge ($A = B = 0$), where $H_L$ is determined by the Einstein field equations. To streamline our notation, we introduce 
\begin{equation}
  f_{k,I}={f_I\over k},\quad\theta_I={v_I-B\over k}. \label{eq:fk_theta}
\end{equation}
In our scheme, $Q_{\rm de}=-Q_{\rm c}=Q$, $\delta Q_{\rm de}=-\delta Q_{\rm c}=\delta Q$, and $f_{\rm de}=-f_{\rm c}=f$, indicating that when $Q$, $\delta Q$, and $f$ are all positive, both energy and momentum transfer from CDM into DE.

Following our previous work \cite{Li:2023fdk}, we directly parameterize the energy transfer perturbation and the momentum transfer potential
\begin{align}
  \delta Q & =C_1\delta_{\rm de}+C_2\delta_{\rm c}+C_3\theta_{\rm de}, \label{eq:parametrized_dQ}\\
  f_k      & =-Q\theta+D_1\theta_{\rm de}+D_2\theta_{\rm c}.\label{eq:parametrized_f}
\end{align}
where $\delta_{\rm de}$($\delta_{\rm c}$) and $\theta_{\rm de}$($\theta_{\rm c}$) represent the fractional density and velocity divergence perturbations for DE (CDM), respectively. Here, $C_1$, $C_2$, $C_3$, $D_1$, and $D_2$ are five time-dependent functions. Different IDE models correspond to different functional forms of $C_1$, $C_2$, $C_3$, $D_1$, and $D_2$. As an example, for the $Q = \beta(a) H \rho_{\rm de}$ model considered in this work, the corresponding coefficient functions are given by $C_1 = D_2 = Q$ and $C_2 = C_3 = D_1 = 0$. Based on this parameterization, we only needs to solve for the model-specific functions in each IDE scenario to obtain the perturbation evolution of DE and CDM. For further related details, see Ref.~\cite{Li:2023fdk}.

However, when calculating the perturbation evolution of DE in IDE models, since vacuum energy ($w = -1$) is not a true geometric background, according to standard linear perturbation theory, DE is treated as a non-adiabatic fluid with negative pressure. When DE interacts with DM, this interaction affects the non-adiabatic pressure perturbations of DE, and may cause the non-adiabatic curvature perturbations to diverge on large scales, leading to what is known as the large-scale instability problem \cite{Majerotto:2009zz,Clemson:2011an}. In other words, cosmological perturbations of DE within IDE models diverge in certain regions of parameter space, potentially leading to the breakdown of IDE cosmology at the perturbation level. To overcome this issue, the parameterized post-Friedmann (PPF) framework \cite{Fang:2008sn,Hu:2008zd} was extended to the IDE models \cite{Li:2014eha,Li:2014cee,Li:2015vla,Li:2023fdk}, known as the ePPF approach. This approach allows for the reliable calculation of cosmological perturbations across the entire parameter space of IDE models. In this work, we employ the ePPF approach to treat the cosmological perturbations (see, e.g., refs.~\cite{Zhang:2017ize,Feng:2018yew}, for applications of the ePPF approach).

\subsection{Cosmological data}\label{sec2.2}

\begin{table}[t]
\caption{Flat priors on the main cosmological parameters constrained in this paper.}
\begin{center}
\renewcommand{\arraystretch}{1.5}
\begin{tabular}{c@{\hspace{0.4cm}}@{\hspace{0.4cm}} c @{\hspace{0.4cm}} c }
\hline\hline
\textbf{Model}       & \textbf{Parameter}       & \textbf{Prior}\\
\hline
$\Lambda$CDM           & $\Omega_{\rm b} h^2$                  & $\mathcal{U}$[0.005\,,\,0.1] \\
                    & $\Omega_{\rm c} h^2$                     & $\mathcal{U}$[0.01\,,\,0.99] \\
                    & $H_0$                                    & $\mathcal{U}$[20\,,\,100] \\
                    & $\tau$                                   & $\mathcal{U}$[0.01\,,\,0.8] \\
                    & $\ln10^{10}A_{\rm s}$                 & $\mathcal{U}$[1.61\,,\,3.91] \\
                    & $n_{\rm s}$                                  & $\mathcal{U}$[0.8\,,\,1.2] \\
\hline
IDE                 & $\beta_0$                                    & $\mathcal{U}$[-6\,,\,2] \\
                    & $\beta_{\rm e}$                                    & $\mathcal{U}$[-2\,,\,8] \\
			
\hline\hline
\end{tabular}
\label{tab1}
\end{center}	
\end{table}

Table~\ref{tab1} provides a summary of the cosmological parameters sampled in different models and the priors applied to them. For the $\Lambda$CDM model analysis, we use six key cosmological parameters: the physical densities of baryons ($\Omega_{\rm b}h^2$) and CDM ($\Omega_{\rm c}h^2$), the Hubble constant ($H_0$), the optical depth ($\tau$), the amplitude of primordial scalar perturbation ($\ln10^{10}A_{\rm s}$), and the scalar spectral index ($n_{\rm s}$). For the IDE models analysis, we include the usual six $\Lambda$CDM parameters and add two dimensionless parameters: the late-time coupling value ($\beta_0$) and the early-time coupling value ($\beta_{\rm e}$). We take $H_0$ as a free parameter instead of the commonly used $\theta_{\rm {MC}}$, because it depends on a standard non-interacting background evolution.

\begin{table*}[!htb]
\setlength\tabcolsep{15pt}{
\renewcommand\arraystretch{2}
\centering
\caption{Fitting results (68.3\% confidence level) in the $\Lambda$CDM and IDE models from the CMB+DESI+DESY5 data. Here, $H_{0}$ is in units of ${\rm km}~{\rm s}^{-1}~{\rm Mpc}^{-1}$.}
\label{tab2}
\begin{tabular}{lcccccc} 
\hline
\hline
Parameter           & $\Lambda$CDM     & IDE1               & IDE2            & IDE3              & IDE4\\ 
\hline
$H_{0}$ & $68.05\pm 0.41$  & $67.04\pm 0.65$              & $67.01\pm 0.67$           & $66.85\pm 0.65$             & $67.10\pm 0.68$              \\
$\Omega_{\mathrm{m}}$  & $0.305\pm 0.005$ & $0.490\pm 0.036$    & $0.335\pm 0.015$          & $0.516^{+0.043}_{-0.037}$  & $0.393^{+0.026}_{-0.034}$    \\
$\beta_0$     & $-$    & $-1.950\pm 0.510$      & $-0.103\pm 0.049$         & $-2.690\pm 0.650$     & $-0.810^{+0.320}_{-0.270}$      \\
$\beta_{\rm e}$    & $-$     & $2.580\pm 0.850$    & $0.004\pm 0.002$        & $4.350^{+1.210}_{-0.970}$      & $0.410\pm 0.140$               \\
\hline
$\ln \mathcal{B}_{ij}$ & $-$              & $3.082$               & $-7.841$        & $3.924$      & $-0.935$    \\
\hline
\hline
\end{tabular}
}
\end{table*}

We compute the theoretical model using the Boltzmann solver {\tt CAMB}, which has been modified to introduce the possibility of interactions between DE and DM\footnote{\url{https://github.com/liaocrane/IDECAMB}.} \cite{Gelman:1992zz,Li:2023fdk}. We perform Markov Chain Monte Carlo (MCMC) \cite{Lewis:2002ah,Lewis:2013hha} 
analyses using the publicly available sampler {\tt Cobaya} \cite{Torrado:2020dgo} and assess the convergence of the MCMC chains using the Gelman-Rubin statistics quantity $R - 1 < 0.02$ \cite{Gelman:1992zz}. The MCMC chains are analyzed using the public package {\tt GetDist} \cite{Lewis:2019xzd}. We use the current observational data to constrain these models and obtain the best-fit values and the $1$--$2\sigma$ confidence level ranges for the parameters of interest \{$H_{0}$, $\Omega_{\mathrm{m}}$, $\beta_0$, $\beta_{\rm e}$\}. Next, we present the observational data used in this work.

\begin{itemize}

\item Cosmic Microwave Background (CMB). Measurements of the Planck CMB temperature anisotropy and polarization power spectra, along with their cross-spectra, as well as the combined ACT and Planck lensing power spectrum. The analysis is based on four key components of CMB likelihoods, as outlined below: 

(i) the power spectra of temperature and polarization anisotropies, $C_{\ell}^{TT}$, $C_{\ell}^{TE}$, and $C_{\ell}^{EE}$, at small scales ($\ell > 30$), are obtained from measurements using the NPIPE PR4 Planck \texttt{CamSpec} likelihood~\cite{Planck:2018vyg,Efstathiou:2019mdh,Rosenberg:2022sdy}; 

(ii) the spectrum of temperature anisotropies, $C_{\ell}^{TT}$, at large scales ($2 \leq \ell \leq 30$), is obtained from measurements using the Planck \texttt{Commander} likelihood~\cite{Planck:2018vyg,Planck:2019nip}; 

(iii) the spectrum of E-mode polarization, $C_{\ell}^{EE}$, at large scales ($2 \leq \ell \leq 30$), is obtained from measurements using the Planck \texttt{SimAll} likelihood~\cite{Planck:2018vyg,Planck:2019nip}; 

(iv) the CMB lensing likelihood, with the latest and most precise data coming from the combination of the NPIPE PR4 Planck CMB lensing reconstruction\footnote{\url{https://github.com/carronj/planck_PR4_lensing}.} \citep{Carron:2022eyg} and Data Release 6 of the ACT\footnote{\url{https://github.com/ACTCollaboration/act_dr6_lenslike}.} \citep{ACT:2023dou}. We denote the combination of these likelihoods as \textbf{``CMB"}.

\item Baryon Acoustic Oscillations (BAO). The DESI BAO data include  tracers from the galaxy, quasar, and the Ly$\alpha$ forest in a redshift range of $0.1 \leq z \leq 4.2$, along with measurements of the transverse comoving distance $D_{\rm {M}}(z)/r_{\rm {d}}$, the Hubble horizon $D_{\rm {H}}(z)/r_{\rm {d}}$, and the angle-averaged distance $D_{\rm {V}}(z)/r_{\rm {d}}$ in 7 distinct redshift bins \citep{DESI:2024uvr,DESI:2024lzq}. Specifically, we use 12 DESI BAO measurements from Ref.~\cite{DESI:2024mwx}.\footnote{\url{https://data.desi.lbl.gov/doc/releases}.} We denote this full dataset as \textbf{``DESI"}.

\item Type Ia Supernovae (SNe). Recently, the DES collaboration released part of the full 5-year dataset, which includes a new, homogeneously selected sample of 1635 photometrically classified SNe (with redshifts in the range $0.1 < z < 1.3$). This is supplemented by 194 low-redshift SNe (with redshifts in the range $0.025 < z < 0.1$), bringing the total number of SNe to 1829 \citep{DES:2024jxu}. We denote this dataset as \textbf{``DESY5"}.\footnote{\url{https://github.com/des-science/DES-SN5YR}.}

\end{itemize}

\section{Results and discussions}\label{sec3}

\begin{figure*}[!htp]
\includegraphics[scale=0.7]{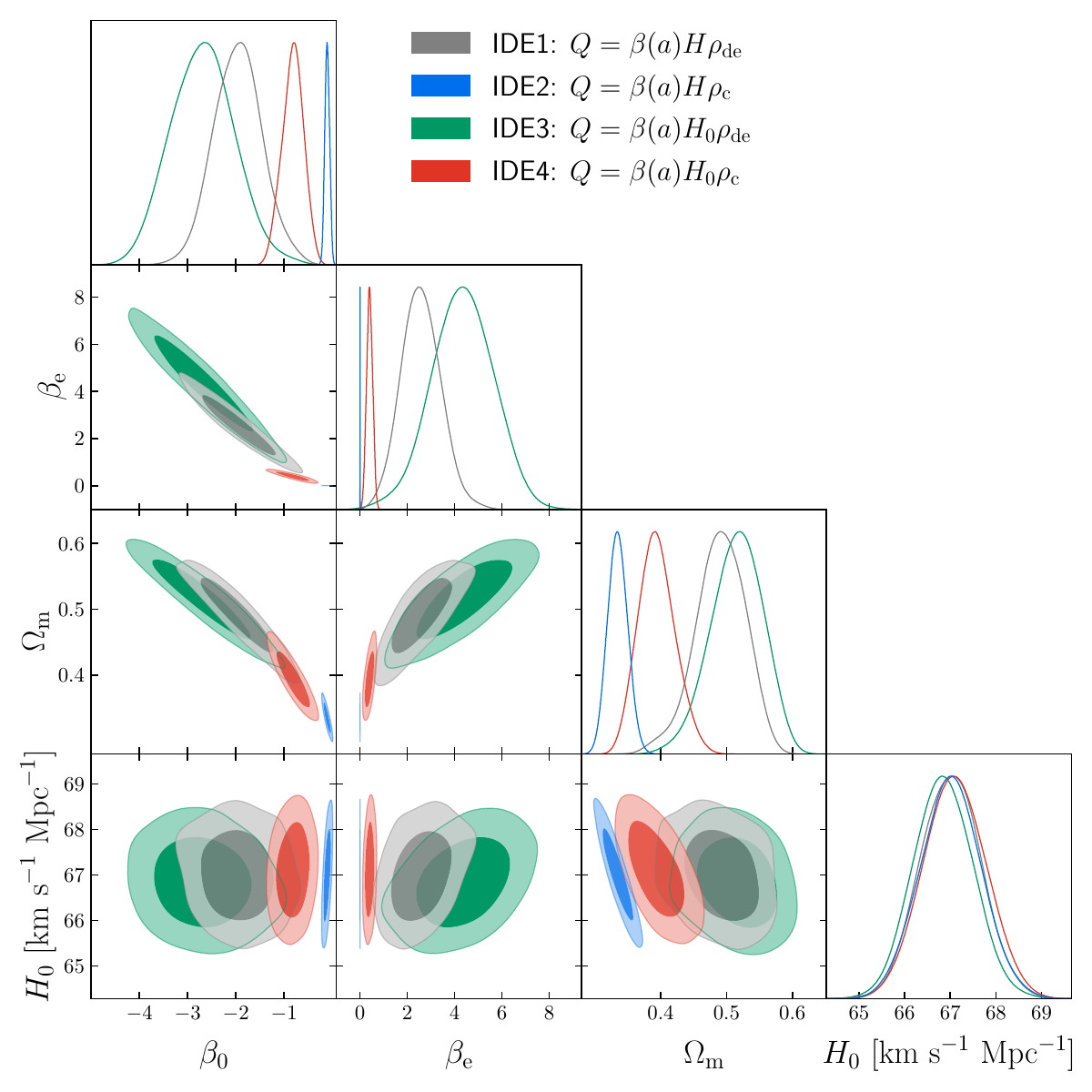}
\centering
\caption{\label{fig1} Constraints on the cosmological parameters using the CMB+DESI+DESY5 data in the IDE1, IDE2, IDE3, and IDE4 models.}
\end{figure*}

\begin{figure*}[htbp]
\resizebox{\textwidth}{!}{%
\includegraphics[]{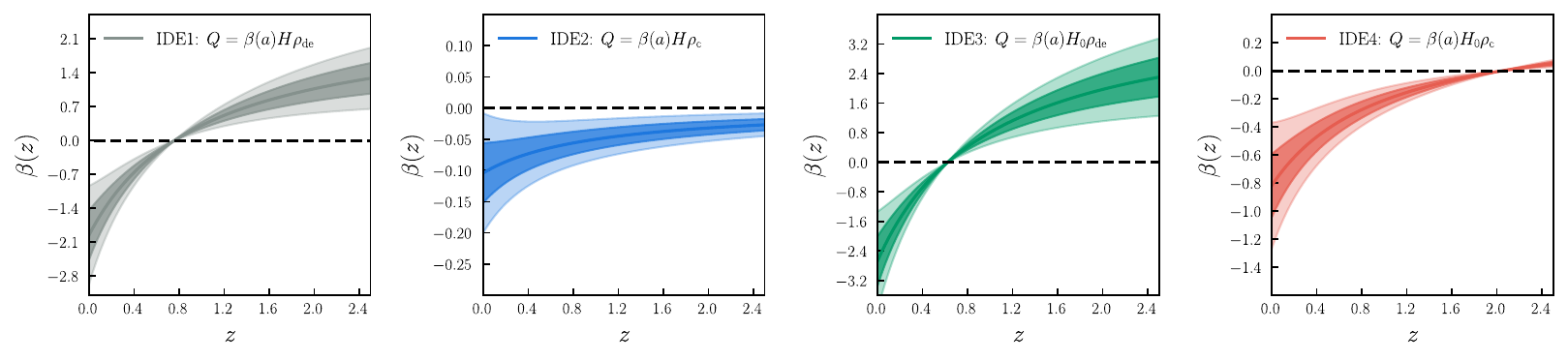}}
\centering
\caption{\label{fig2} Reconstructed evolutionary history of $\beta(z)$ at $1\sigma$ and $2\sigma$ confidence levels in the IDE1, IDE2, IDE3, and IDE4 models. The black dashed line in each plot represents the non-interacting line $\beta(z)=0$.}
\end{figure*}

\begin{figure*}[htbp]
\includegraphics[scale=0.65]{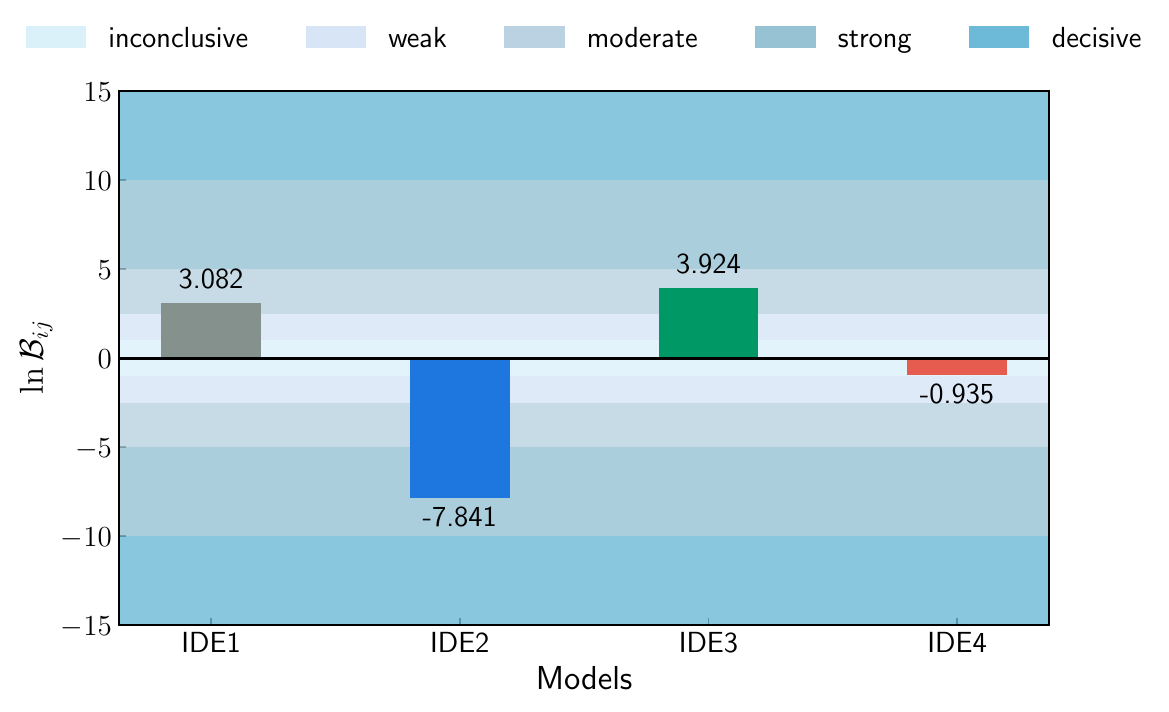}
\centering
\caption{\label{fig3} Comparison of the Bayesian evidence for the IDE models and the $\Lambda$CDM model. The Bayes factor $\ln \mathcal{B}_{ij}$ (where $i$ = IDE, $j$ = $\Lambda$CDM) and its strength according to the Jeffreys scale are used to assess the preference between models, where a positive value indicates a preference for the IDE model.}
\end{figure*}

In this section, we shall report the constraint results of the cosmological parameters. We consider the four IDE models to perform a cosmological analysis using current observational data, including DESI BAO, CMB, and DESY5 SN data. We show the $1\sigma$ and $2\sigma$ posterior distribution contours for various cosmological parameters in the four IDE models, as shown in Fig.~\ref{fig1}. The $1\sigma$ errors for the marginalized parameter constraints are summarized in Table~\ref{tab2}. We reconstructed the evolution history of $\beta(z)$ at the $1\sigma$ and $2\sigma$ confidence levels for the four IDE models, as shown in Fig.~\ref{fig2}. Finally, we compared $\ln \mathcal{B}_{ij}$ between the $\Lambda$CDM and the IDE models using the current observational data, as shown in Fig.~\ref{fig3}.

In Fig.~\ref{fig1}, we show the triangular plot of the constraint results for the four IDE models using the CMB+DESI+DESY5 data, here we also explicitly present these results:

\begin{itemize}

\item For the IDE1 model, we have $\beta_{0}=-1.950\pm 0.510$, $\beta_{\rm e}=2.580\pm 0.850$, $\Omega_{\rm m}=0.490\pm 0.036$, and $H_0=67.04\pm 0.65~\rm km~s^{-1}~Mpc^{-1}$.
\item For the IDE2 model, we have $\beta_{0}=-0.103\pm 0.049$, $\beta_{\rm e}=0.004\pm 0.002$, $\Omega_{\rm m}=0.335\pm 0.015$, and $H_0=67.01\pm 0.67~\rm km~s^{-1}~Mpc^{-1}$.
\item For the IDE3 model, we have $\beta_{0}=-2.690\pm 0.650$, $\beta_{\rm e}=4.350^{+1.210}_{-0.970}$, $\Omega_{\rm m}=0.516^{+0.043}_{-0.037}$, and $H_0=66.85\pm 0.65~\rm km~s^{-1}~Mpc^{-1}$.
\item For the IDE4 model, we have $\beta_{0}=-0.810^{+0.320}_{-0.270}$, $\beta_{\rm e}=0.410\pm 0.140$, $\Omega_{\rm m}=0.393^{+0.026}_{-0.034}$, and $H_0=67.10\pm 0.68~\rm km~s^{-1}~Mpc^{-1}$.

\end{itemize}

For these results, we find that the statistical significances for $\beta_0 < 0$ and $\beta_{\rm e} > 0$ are at 3$\sigma$, 2$\sigma$, 4.1$\sigma$, and 2.5$\sigma$ levels for the IDE1, IDE2, IDE3, and IDE4 models, respectively. Additionally, an anti-correlation between $\beta_0$ and $\beta_{\rm e}$ in the $\beta_0$--$\beta_{\rm e}$ plane is explicitly shown in Fig~\ref{fig1}. It is noted that such an anti-correlation was first observed using early data in a previous study \cite{Li:2011ga}. In ref.~\cite{Guo:2017deu}, a stronger anti-correlation was found, with $\beta_0$ and $\beta_{\rm e}$ exhibiting a linear relationship for all IDE models. However, our latest results indicate that the degree of anti-correlation remains strong in IDE1 and IDE3 models, slightly weak in the IDE4 model, and very weak in IDE2 model. We find that IDE2 model gives the best constraints on the parameters $\beta_{\rm e}$ and $\beta_0$, followed by IDE4 model, while IDE1 and IDE3 models give comparable constraint results. In the IDE2 model, the constraints on $\beta_{\rm e}$ are significantly tighter compared to the $\beta_0$. This is because, in the early universe, both $H$ and $\rho_{\rm c}$ have relatively high values, and the parameter $\beta_{\rm e}$, which characterizes the early universe, must therefore be very small. As a probe of the early universe, CMB data can provide relatively stringent constraints on the early-time coupling value $\beta_{\rm e}$, whereas the constraints on the late-time coupling value $\beta_0$ are somewhat weaker.

As shown in the $H_0$--$\Omega_{\rm m}$ plane of Fig~\ref{fig1}, we find that for the four IDE models, $H_0$ and $\Omega_{\rm m}$ are anti-correlated, as the case in the base $\Lambda$CDM model. For the four IDE models, the central value of $H_0$ is generally consistent and slightly lower, while that of $\Omega_{\rm m}$ is higher, compared to the $\Lambda$CDM model. In addition, the central value of $\Omega_{\rm m}$ is higher for IDE1 and IDE3 models than for IDE2 and IDE4 models, with the range of $\Omega_{\rm m}$ being significantly larger. This is because $\Omega_{\rm m}$ is influenced by the parameter $\beta_0$, whereas $H_0$ is only minimally affected by $\beta_0$. As shown in the $\beta_0$--$\Omega_{\rm m}$ and $\beta_0$--$H_0$ planes of Fig~\ref{fig1}, $\Omega_{\rm m}$ is inversely correlated with $\beta_0$ and shows a stronger correlation in IDE1 and IDE3, while $H_0$ exhibits virtually no correlation with $\beta_0$.

In Fig~\ref{fig2}, we present the reconstructed evolution of $\beta(z)$ based on the best-fit values of $\beta_{\rm e}$ and $\beta_0$, along with the 1$\sigma$ and 2$\sigma$ confidence levels. In the previous study \cite{Guo:2017deu}, $\beta(z)$ crosses the non-interacting line $\beta(z) = 0$ for all IDE models. However, based on the results of this work, we find that for the IDE1, IDE3, and IDE4 models, $\beta(z)>0$ at early times and $\beta(z)<0$ at late times, with $\beta(z)$ crossing the non-interacting line $\beta(z)=0$ during cosmic evolution at a confidence level exceeding 2$\sigma$. For the IDE2 model, $\beta(z)$ remains negative throughout and does not cross $\beta(z) = 0$ at the 2$\sigma$ confidence level, implying that the interaction between DE and DM does not experience a sign change during cosmic evolution. This also highlights the importance of considering various IDE models to investigate whether a sign change interaction between DE and DM exists during cosmic evolution. The crossing happens at a redshift around 0.5--0.7 for the IDE1 and IDE3 models, and at $z\sim 2.0$ for the IDE4 model. We further find that when DM dominates the universe, energy transfer occurs from DM to DE, and when DE dominates the universe, energy transfer occurs from DE to DM, for the IDE1 and IDE3 models. The results of this study lead to conclusions that are partially similar to those in previous studies \cite{Li:2011ga,Guo:2017deu}. Our results exclude the possibility of a sign change in the interaction within the IDE2 model at the 2$\sigma$ confidence level and provide stronger evidence for a sign-changeable interaction in the IDE1, IDE3, and IDE4 models at the 3$\sigma$, 4.1$\sigma$, and 2.5$\sigma$ confidence levels, respectively.
 
Finally, we use the Bayesian evidence selection criterion to choose the preferred IDE models over the $\Lambda$CDM model based on the current observational data. To compute the Bayesian evidence of the models, we employ the publicly available code {\tt MCEvidence}\footnote{\url{https://github.com/yabebalFantaye/MCEvidence}.} \cite{Heavens:2017afc,Heavens:2017hkr}. The Bayesian evidence $Z$ is defined as
\begin{equation}
Z = \int_{\Omega} P(D|\bm{\theta},M)P(\bm{\theta}|M)P(M)\ {\rm d}\bm{\theta},
\label{eq: lnZ}
\end{equation}
where $P(D|\bm{\theta},M)$ is the likelihood of the data $D$ given the parameters $\bm{\theta}$ and the model $M$, $P(\bm{\theta}|M)$ is the prior probability of $\bm{\theta}$ given $M$, and $P(M)$ is the prior of $M$. Then we calculate the Bayes factor $\ln \mathcal{B}_{ij} = \ln Z_i - \ln Z_j$ in logarithmic space, where $Z_i$ and $Z_j$ are Bayesian evidence of two models.

The strength of model preference is typically assessed using the Jeffreys scale \citep{Kass:1995loi,Trotta:2008qt}.  According to this scale: if $\left|\ln \mathcal{B}_{ij}\right|<1$, the evidence is inconclusive; $1\le\left|\ln \mathcal{B}_{ij}\right|<2.5$ represents weak evidence; $2.5\le\left|\ln \mathcal{B}_{ij}\right|<5$ is moderate; $5\le\left|\ln \mathcal{B}_{ij}\right|<10$ is strong; and if $\left|\ln \mathcal{B}_{ij}\right|\ge 10$, the evidence is decisive.

In Fig~\ref{fig3}, we show a comparative analysis of $\ln \mathcal{B}_{ij}$ for the four IDE models relative to the $\Lambda$CDM model, using the current observational data. Here, $i$ denotes the IDE model and $j$ denotes the $\Lambda$CDM model. The positive value indicates a preference for the IDE models over the $\Lambda$CDM model, while negative values indicate a preference for the $\Lambda$CDM model. The $\ln \mathcal{B}_{ij}$ values for the four IDE models relative to the $\Lambda$CDM model, based on CMB+DESI+DESY5 data, are as follows: $3.082$ for IDE1, $-7.841$ for IDE2, $3.924$ for IDE3, and $-0.935$ for IDE4. Our results indicate that the IDE2 model is strongly disfavored, the IDE4 model is inconclusive, and the IDE1 and IDE3 models are moderately preferred over the $\Lambda$CDM model.

\section{Conclusion}\label{sec4}

Ever since the proposal of interaction between DE and DM, these cosmological scenarios have attracted widespread attention in the scientific community, particularly the phenomenological IDE models with unidirectional energy flow (the energy transfer can occur either from DE to DM or from DM to DE). However, according to theoretical and observational findings, the direction of energy flow may reverse in the late-time universe, but this conclusion depends on the underlying interaction model and observational data. Therefore, this is also an interesting and worthwhile issue to investigate in the context of IDE scenarios. 

In this work, we aim to investigate whether the interaction between DE and DM changes sign during cosmic evolution, using the DESI BAO, CMB, and DESY5 SN data. In order to obtain comprehensive results, we investigate four IDE models, i.e., the IDE1 model with $Q = \beta(a) H \rho_{\rm de}$, the IDE2 model with $Q = \beta(a) H \rho_{\rm c}$, the IDE3 model with $Q = \beta(a) H_0 \rho_{\rm de}$, and the IDE4 model with $Q = \beta(a) H_0 \rho_{\rm c}$. We also perform model selection using Bayesian evidence to determine the preferred form of the IDE models from the current observational data compared to the $\Lambda$CDM model.

Our findings indicate $\beta_0 < 0$ and $\beta_{\rm e} > 0$ at a confidence level of above 2$\sigma$ for all the IDE models. We observe an anti-correlation between $\beta_0$ and $\beta_{\rm e}$, which is strongest in the IDE1 and IDE3 models, weaker in the IDE4 model, and weakest in the IDE2 model. For the IDE models (except for the IDE2 model), $\beta(z) > 0$ at early times and $\beta(z) < 0$ at late times, with the coupling $\beta(z)$ crossing the non-interacting line $\beta(z) = 0$ during cosmological evolution at a confidence level exceeding 2$\sigma$. For the IDE2 model, $\beta(z)$ remains negative throughout and does not cross $\beta(z) = 0$ at the 2$\sigma$ confidence level. We further find that the energy transfer occurs from DM to DE when DM dominates the universe, and from DE to DM when DE dominates, for the IDE1 and IDE3 models. These results lead to conclusions similar to those in previous studies \cite{Li:2011ga,Guo:2017deu}. However, our results exclude the possibility of a sign change in the interaction within the IDE2 model at the 2$\sigma$ confidence level, while providing stronger evidence for the occurrence of a sign change in the other IDE models. Specifically, in the IDE1 and IDE3 models, the existence of a sign-changeable interaction is supported by the current data at the 3$\sigma$ and 4.1$\sigma$ confidence levels, respectively. Additionally, the Bayesian evidence indicates that the IDE2 model is strongly disfavored, the IDE4 model is inconclusive, and the IDE1 and IDE3 models are moderately preferred over the $\Lambda$CDM model, especially for the IDE3 model with $\ln \mathcal{B}_{ij}$ = $3.924$. We can conclude that the IDE3 model has an advantage over the $\Lambda$CDM model in fitting the current observational data. In the coming years, the complete DESI dataset, combined with more precise late-universe data from Euclid \cite{Euclid:2024yrr} and Large Synoptic Survey Telescope \cite{LSST:2008ijt}, as well as future CMB observations from the CMB-S4 \cite{CMB-S4:2016ple}, could be utilized to investigate sign-changeable IDE models, potentially uncovering more unexpected and intriguing results.

\section*{Acknowledgments}
We thank Sheng-Han Zhou, Yue Shao, Ji-Guo Zhang, and Hai-Li Li for their helpful discussions. This work was supported by the National SKA Program of China (Grants Nos. 2022SKA0110200 and 2022SKA0110203), the National Natural Science Foundation of China (Grants Nos. 12533001, 12575049, and 12473001), the China Manned Space Program (Grant No. CMS-CSST-2025-A02), and the National 111 Project (Grant No. B16009).  

\bibliography{main}

\end{document}